\documentclass[conference]{IEEEtran}
\usepackage{comment}
\usepackage{amssymb,mathtools}
\usepackage{nccmath}
\usepackage{adjustbox}

\usepackage{graphicx}

\usepackage{circledsteps}

\usepackage{subfig}

\usepackage{diagbox}
\usepackage{multirow}

\ifCLASSINFOpdf
\else
 \fi

\usepackage[utf8]{inputenc}
\usepackage[english]{babel}
\usepackage{fancyhdr}

\begin{document}

\title{Too Expensive to Attack: A Joint Defense Framework to Mitigate Distributed Attacks for the Internet of Things Grid}
\author{\IEEEauthorblockN{Jianhua Li}
\IEEEauthorblockA{\textit{UniCloud Australia}\\
Melbourne, Australia \\
jackleeml@hotmail.com}
\and
\IEEEauthorblockN{Ximeng Liu}
\IEEEauthorblockA{\textit{Fuzhou University}\\
Fuzhou, China \\
snbnix@gmail.com}
\and
\IEEEauthorblockN{Jiong Jin}
\IEEEauthorblockA{\textit{Swinburne University of Technology}\\
Melbourne, Australia \\
jiongjin@swin.edu.au}
\and
\IEEEauthorblockN{Shui Yu}
\IEEEauthorblockA{\textit{University of Technology Sydney}\\
Sydney, Australia \\
shui.yu@uts.edu.au}
}
\maketitle

\begin{abstract}
The distributed denial of service (DDoS) attack is detrimental to businesses and individuals as we are heavily relying on the Internet. Due to remarkable profits, crackers favor DDoS as cybersecurity weapons in attacking servers, computers, IoT devices, and even the entire Internet. Many current detection and mitigation solutions concentrate on specific technologies in combating DDoS, whereas the attacking expense and the cross-defender collaboration have not drawn enough attention. Under this circumstance, we revisit the DDoS attack and defense in terms of attacking cost and populations of both parties, proposing a joint defense framework to incur higher attacking expense in a grid of Internet service providers (ISPs), businesses, individuals, and third-party organizations (IoT Grid). Meanwhile, the defender's cost does not grow much during combats. The skyrocket of attacking expense discourages profit-driven attackers from launching further attacks effectively. The quantitative evaluation and experimental assessment reinforce the effectiveness of our framework.
\end{abstract}


\IEEEpeerreviewmaketitle

\section{Introduction}

\IEEEPARstart{I}{t} is unthinkable for many people living without the Internet \cite{li2015ehopes}. Going offline for a few hours may bring very negative feelings, including business losses, social life ruins, family concerns, and low productivity. More interestingly, the Internet of things (IoT) is evolving to reshape the future of businesses and individuals \cite{forestiero2017iot}. The IoT tends to be an integration of people, data, processes, and devices, helping us to live smarter and work smarter. The stakeholders of IoT include businesses, individuals, ISPs, and third-party organizations, composing a family of IoT grids for different communities on the Internet \cite{liu2018umbrella}. Due to the significant impact, Internet availability and data accessibility are often a fundamental right to achieving the vision of our future.

Unfortunately, this fundamental right often suffers from viruses, malware, spyware, ransomware, phishing, and various distributed denial of service (DDoS) attacks. Crackers favor distributed attacks because of their low cost, high returns, and easiness of implementation. Initially, numerous IoT devices are more vulnerable and much easier to compromise. Meanwhile, attackers can use compromised IoT devices as bots, attacking a variety of potential victims, such as servers, IoT devices, bandwidth, and even the Internet, for financial benefits. Also, a defender has limited capability to trace and catch these attackers as they hide behind the Internet. In consequence, DDoS attacks seem to be a nightmare for the Internet that are malicious attempts to disrupt the regular traffic and resource scheduling for defenders.

\begin{figure}[t!]
\centering
\includegraphics[width=1\columnwidth]{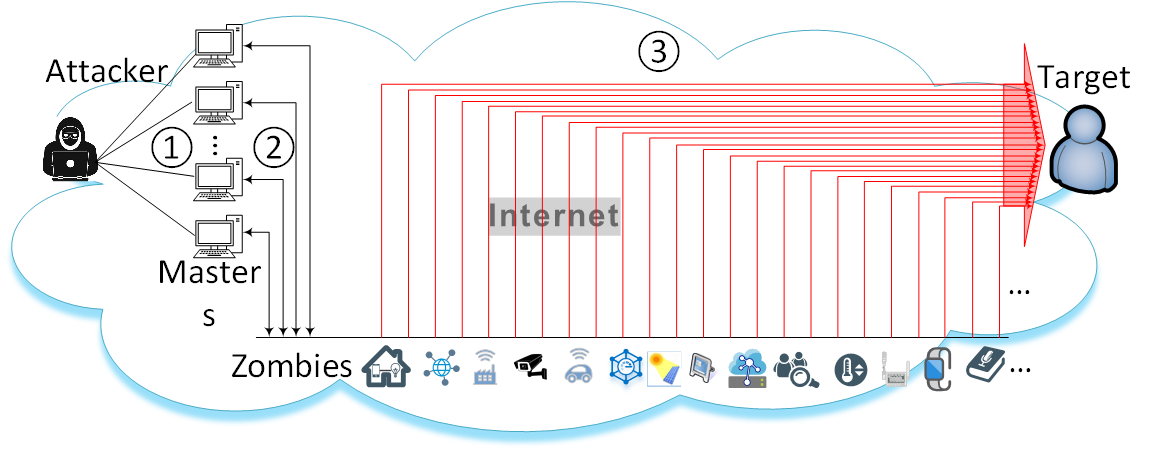}
\caption{Botnets-based attack: an attacker builds a group of masters (step \raisebox{.5pt}{\textcircled{\raisebox{-.9pt} {1}}}) from that to capture and control zombies (step \raisebox{.5pt}{\textcircled{\raisebox{-.9pt} {2}}}), then using distributed zombies to attack a user (step \raisebox{.5pt}{\textcircled{\raisebox{-.9pt} {3}}}). Such zombies are computing nodes locates in distributed networks of users. Due to a lack of collaboration between users, a user's network becomes an attacking source while another user's network is the victim.}
\label{f1}
\end{figure}

As an essential component of a botnet \cite{kolias2017ddos}, a bot is a lightweight software instance running on a device automatically and autonomously. Such devices turn to be networked zombies under the uniform control of an attacker, able to retrieve and carry out instructions for illicit hacking purposes, as shown in Fig. 1. The malicious code targeting remote victims from botnets seems to be regular business data in the local network. It is hard to detect bots unless they actively attack a local entity. It is even more challenging to charge the attacker as they conduct illegal activities (e.g., fraud) from devices of other owners. In a DDoS attack, attackers manage to overwhelm the target or its surrounding infrastructure with a flood of data flow at a time, causing resource starvation for regular business usages. To this end, the IoT devices represent a large volume of exploited machines that generate an unexpected traffic jam clogging up the Internet, preventing regular data flows from arriving at their destination. According to Amazon Web Services (AWS), the worst DDoS attacked bandwidth of more than 2.3 Tbps in mid-February 2020 \cite{Cimpanu2020aws}.





\begin{table}[t!]
\centering
\caption{Reported price of botnet for hire services}
\label{T1}
\begin{adjustbox}{max width=1\columnwidth}
\begin{tabular}{|c|c|c|c|}
\hline
Botnet             & Bot Types           & Population & Rental Price (\$) \\ \hline
Botnet-Canada      & Computers           & 1000       & 270               \\ \hline
Botnet-the U. S.   & Computers           & 1000       & 180               \\ \hline
Botnet-the U. K.   & Computers           & 1000       & 240               \\ \hline
Botnet-France      & Computers           & 1000       & 200               \\ \hline
Boy Webcam         & Hacked IIoT device  & 100        & 1                 \\ \hline
Girl Webcam        & Hacked IIoT device  & 100        & 100               \\ \hline
Remote controlller & Administration tool & 1          & 40                \\ \hline
\end{tabular}
\end{adjustbox}
\end{table}

Although the motivation varies from business competition to cybercriminals, both DDoS attacking and defending have been becoming highly profitable businesses. In particular, DDoS for hire services are available in many countries. Table I presents the price, location, and types of hacked devices for hire in the market. Kaspersky Lab's experts reported that arranging an attack costs only \$7 per hour, using a cloud-based botnet of 1000 desktops. Meanwhile, the average price for an attack is \$25 per hour such that an attacker can make an \$18 profit per hour for one attacking, approximately. Surely, attackers may demand a ransom from a user for not attacking an object, leading to a remarkable return of one attack \cite{ismail2017criminal}.

With such profits in mind, we develop a fog-based joint defense framework to enlarge the attack cost \cite{Somani2017cost} in the IoT grid, thus discouraging DDoS attackers. Due to its proximity to attacking sources, fog can quickly locate malicious codes and perform filtering at an earlier stage with comparison to individual mitigation at the victim-end. Our contribution is fourfold:

\begin{itemize}
    \item We investigate the attacking expenses of DDoS attacks and found botnet expense is the primary factor to enlarge the cost of attacking.
    \item We advocate joint-mitigation for defenders to handle distributed attacks coordinated by an attacker. Compared with individual defense, the joint defense approach sets the front line to the attacking source, potentially captures the zombies and the attack masters, and mitigates attacks at the IoT edge.
    \item Our approach benefits from existing mitigation infrastructure, maintaining low expense at the defense-side. However, it incurs a high cost of the attack at the attacker-side during the combat. In consequence, it prevents profit-driven attackers in the market.
    \item We compare the attacking expense between individual and the joint defense framework to showcase the advantage. Meanwhile, we implement the framework in our testbed, illustrating the quality of experience (QoE) improvement with the increasing number of defenders.
\end{itemize}

The road map of this paper is as follows. Section II studies related work, and Section III presents the attack model and assumptions. We explore the attacking cost and population of both parties during combat in Section IV and propose our joint defense framework in Section V. Then, Section VI scrutinizes the cost evaluation and experimental assessment. In Section VII, we review the benefits and possible challenges and briefly summarize the works studied in this paper. Finally, we conclude the paper in Section VIII.

\section{Related Work}
As explored in Table II, DDoS attacks continuously grow in sizes and frequencies, causing severe consequences in many aspects. Therefore, DDoS attacks and mitigation solutions have been a hot topic in the digital security research community for many years. Scholars put forward a variety of technological advancements in fighting DDoS attacks, encompassing software-defined networking (SDN), blockchain, network function virtualization (NFV), machine learning, and big data, etc. 

In \cite{kalkan2017filtering}, Kalkan \textit{et al.} reviewed filtering-based defense mechanisms against DDoS, pinpointing the complexity to combat coordinated DDoS attacks. Due to high dynamics in the IoT, the authors claimed that a highly effective filtering scheme might quickly become obsolete. Furthermore, they concluded machine learning, NFV, SDN, and cloud computing could offer a promising solution to beat DDoS attacks.

\begin{table}[t!]
\caption{Reported DDoS attacks}
\centering
\label{T2}
\begin{adjustbox}{max width=1\columnwidth}
\begin{tabular}{|c|c|c|c|}
\hline
Time      & Attack Scale & Reporter   & Main Methods            \\ \hline
15-Feb-20 & 2.3 Tbps     & AWS        & UDP flooding            \\ \hline
Jan-19    & 396 Gbps     & Imperva    & SYNC flooding           \\ \hline
Mar-18    & 1.7 Tbps     & NETSCOUT   & Memcached amplification \\ \hline
28-Feb-18 & 1.35 Tbps    & GitHub     & Git attack              \\ \hline
Oct-16    & 1.2 Tbps     & Dyn        & DNS flooding            \\ \hline
Sep-16    & 1 Tbps       & Scmedia    & TCP Syn, Ack            \\ \hline
Jul-15    &              & GitHub     & HTTP flooding           \\ \hline
Nov-14    & 500 Gbps     & Forbes     &                         \\ \hline
Feb-14    & 400 Gbps     & Cloudflare & NTP                     \\ \hline
Jul-13    & 300 Gbps     & Cloudflare & DNS flooding            \\ \hline
\end{tabular}
\end{adjustbox}
\end{table}


Thanks to the programmability, global view, and centralized management, SDN could remove the heavy reliance on other systems fighting against DDoS attacks. Sahay \textit{et al.} proposed an SDN-based autonomic mitigation framework in \cite{SAHAY2017aroma}, offering broad protection coverage, easy deployment, dynamic mitigation, and automation. Their scheme allows ISPs to defend against attacks on behalf of customers, which in turn reduces the collateral damage in the ISP network. Even so, SDN itself introduces new DDoS attack surfaces like flow table overloading. Bhushan \textit{et al.} applied the queuing theory to share flow tables and meanwhile keep minimal involvement of the SDN controller in  \cite{bhushan2019distributed}. 

El Houda \textit{et al.} combined SDN with blockchain technologies in DDoS mitigation for smart environments \cite{ElHouda2019blockchain}, using smart contracts to transfer attack information among distributed networks. Yan \textit{et al.} proposed the SDN-empowered MLDMF that is a hierarchical framework spanning cloud, fog, and edge to mitigate DDoS attacks in \cite{Yan2018IIoT}. According to this proposal, the edge focuses on access control, fog achieves intrusion detection, while the cloud accomplishes DDoS detection.

Machine learning has its advantages of automation and accuracy in analyzing a large volume of traffics on the Internet. In this regard, Ko \textit{et al.} claimed using deep learning to mitigate DDoS within the ISP domain in \cite{ko2020feature}. This approach delivers a more efficient solution as it gets rid of the data detection site (e.g., auxiliary server, data scrubbing centers) at the client's side. Nonetheless, it will probably incur other security issues like privacy.  Also, Simpson \textit{et al.} designed a per-host DDoS mitigation approach with reinforcement learning in \cite{Simpson2020reinforcementlearning}, aiming at classifying malicious traffic without investigating protocols. The authors showcased a significant TCP goodput improvement compared with other models. Li \textit{et al.} developed an attacker-centric and federated learning empowered framework to reduce mitigation time and improve detection accuracy in \cite{li2020fleam}, where the mitigation delay is about 72\% lower while the accuracy is 47\% greater on average than previous solutions.

Note that big data has benefits in recognition of malicious traffic with improved efficiency. For example, Zhou \textit{et al.} proposed an online traffic monitoring framework for performance monitoring and DDOS attack detection in \cite{zhou2018online}. Compared with offline data analysis frameworks, the framework has the benefits of quick response to attacks and high accuracy in identifying various flooding attacks.

Zhou \textit{et al.} proposed a fog computing framework in handing DDoS threats \cite{ZHOU2019fog} for an IIoT system. The framework has three tiers, i.e., field control, local control, and cloud control. Firewalls sit between filed control and local control layer, protecting filed devices against attacks from upper tiers. Servers in the local control layer perform DDoS detection and report the result to cloud servers for central consolidation and analysis. The authors intended to allocate traffic flow analysis workloads to multiple servers in a flexible manner by leveraging the NFV technique. 

\section{Attack Model and Assumptions}
Different from the above work that focuses on combating DDoS in a specific solution, we work on a joint defense framework that contains these technologies for value-added services. By reducing potential attacking profits, we discourage attackers from launching DDoS attacks in the IoT grid. The primary strategy for the low-cost attack is using zombies to concentrate the superior forces and defeat each defender, individually and respectively. Without zombies, an attacker has to build attacking agents in private infrastructure, which is much more costly and easy to catch. In this regard, the joint defense is beneficial to every defender.

Fig. 2 shows our attack model, where each customer performs prevention, detection, and mitigation as usual \cite{bawany2017ddos}. Benign traffic and malicious data flow compose the total number of arrivals to a victim for a given time in an attack.  Various methods of fighting with DDoS attacks may include challenge-response  \cite{venkatesan2016moving}, reputation-based access \cite{otung2020distributed}, resource limits \cite{javaid2018mitigating}, anomaly detection \cite{doshi2018machine}, source tracing \cite{chen2016dispersing}, filtering, resource scaling \cite{yu2013can}, etc. It is worth mentioning that the goal of our article is to strengthen cooperation against attacks from the perspective of attacking expense. We, therefore, do not focus on such a specific method, as our approach allows each customer to employ individual technologies.

In essence, the attack and defense is a game of resource competition \cite{yu2013can}. If an attacker has sufficient resources to incur more resource depletion to a defender than the defender can supply, then the attacker will win, and vice versa. 
\begin{equation}
    Attack_{win}=  
    \begin{cases}
       1,  & \text{if } R_{consume} > R_{supply}\\
       0,  & \text{if } R_{consume} \leq R_{supply}  
    \end{cases}
\end{equation}
where $R$ stands for resources.
With zombies, an attacker can turn one user's resources into attacking resources from which to attack another user. To this end, we assume that the decrements of attacking resources signal the increment of defending resources in our joint defense system. 

\begin{figure}[t!]
\centering
\includegraphics[width=\columnwidth]{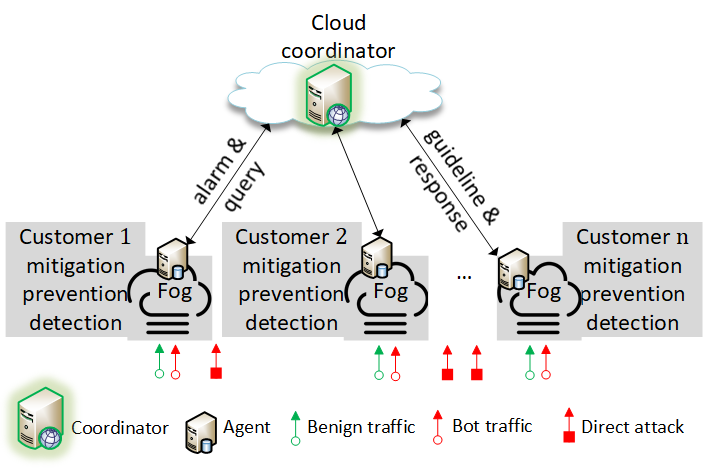}
\caption{Joint defense in fog: Agents send query and alarm messages to the coordinator and peers, while the coordinator propagates alarms and guidelines, responding to queries. Agents of each fog network instruct defending nodes for collaborative defense.}
\label{f2}
\end{figure}

To facilitate the cross-defender collaboration, each customer who participates in the scheme needs to install a DDoS defending agent. The agent can send alarm messages and queries to peers and the cloud coordinator that has a global view of the defending system. The coordinator propagates alarm messages to each agent as well as informs each agent of actionable knowledge of potential bots. Grounded in guidelines and updates retrieved from the coordinator, each participant can take actions to prevent, detect, and mitigate bots in collateral networks. Besides, we assume these messages to be delivered securely and reliably. 

Last, we assume that attackers are profit-driven. They will stop attacking a customer if it is not profitable. Furthermore, IoT networks connect to the cloud via fog networks. Defenders can perform victim-end, core, source-end detection, and mitigation when possible.  

\section{DDoS Attack and Joint Defense}

To fight against distributed attacks, we advocate joint defense of businesses, individuals, ISPs, and third-party organizations, as illustrated in Fig. 3. Under this circumstance, defenders force the attacking line to exposure when malicious codes traverse the defense zone. Once identified, defenders can limit such malicious traffic and track the attacking masters and the attacker at the edge of the IoT grid. This approach incurs a high cost of attacking, thus potentially preventing future attacks on participants. 

\subsection{Attacking Cost}
There are three types of traffics, including benign data, bots flow, and direct attacking code (DAC) from the attacker's network. During an attack, the total resource consumed by these data is:

\begin{equation}
  \begin{split}
    Resource_{attack} &= Resource_{benign} + Resource_{bot} \\
    & + Resource_{direct} 
    \end{split}
\end{equation}

\begin{figure}[t!]
\centering
\includegraphics[width=\columnwidth]{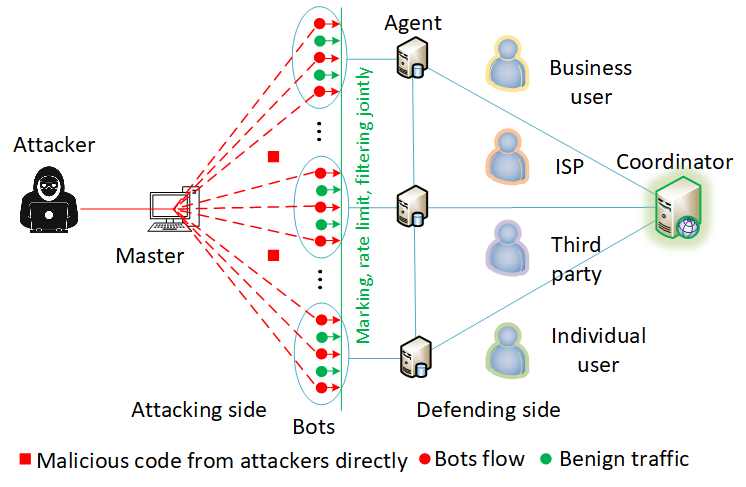}
\caption{Joint attacking and defending.}
\label{f3}
\end{figure}

Either of them can cause resource consumption but at different costs. The benign traffic is free of charge, and the bot traffic is cheap, whereas the DAC is more costly. In particular, a network of malicious nodes is notorious and easy to identify via reputation-based detection. The defender party can take a quick response to blacklist such notorious nodes, leading to less attacking time and higher attacking cost.

\begin{equation}
  \begin{split}
    Cost_{attack} & = Cost_{benign} + Cost_{bot} + Cost_{direct} \\
               & = Cost_{bot} + Cost_{direct}\\ 
  \end{split}
\end{equation}

Although the cost of benign traffic is almost nothing, such traffic is usually regular and statistically knowable. Defenders often reserve sufficient resources in processing these business data. Besides, the DAC is often too expensive to attack an object. Quite often, attackers use these computing nodes as masters for the management of zombies.  As a result, an attacker usually relies on botnets as low-cost attacking sources.

\begin{equation}
    Cost_{bot} = Expense_{setup} + Expense_{rental}
\end{equation}

The setup expense per bot is relatively a fixed value around \$1 each. The rental cost varies according to the population of a botnet. According to \cite{cimpanu2016bot}, renting around 400,000 bots costs between \$20,000-\$30,000 for 2 weeks. If a bot is identified and mitigated, the attacker will lose the offensive power to the defender, so the mitigation response time (MRT) is a determining factor to the cost of per bot during combat. To illustrate, if the MRT is more than two weeks, the per bot cost is only \$0.05 to 0.075 for two weeks. If the MRT is one hour, the per active bot expense ($PABE$) is \$(0.05~0.075)*24*14 = \$16.8 to 25.2 for two weeks. The MRT is up to the defensive power of a defender's system. 

\begin{figure}[t!]
\centering
\includegraphics[width=0.9\columnwidth]{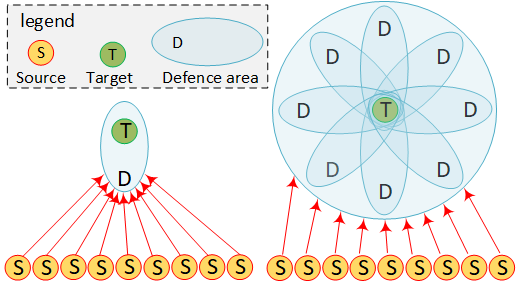}
\subfloat[\label{a} individual]{\hspace{.5\linewidth}}
\subfloat[\label{b} joint]{\hspace{.5\linewidth}}
\caption{Attacking and defending: Individual vs. joint.}
\label{f4}
\end{figure}

According to Equation (1), the attacker has to incur enough attacking traffic to suppress resource supply for a given time. In consequence, the attacker must keep a certain number of active bots as well as direct attacking points $Num_{actv}$ to continuously suppress the victim. In this regard, the botnet expense is:

\begin{equation}
    Expense_{botnets} = Num_{actv}*PABE
\end{equation}

In a collaborative defense environment, the specific value of active bots is up to the number of defending units $Num_{unit}$. The correlation obeys the Lokta-Volterra Law:

\begin{equation}
    \begin{cases}
       \frac{dNum_{unit}}{dt} &= \alpha Num_{unit} - \beta Num_{unit} Num_{actv}   \\
       \frac{dNum_{actv}}{dt} &= \delta Num_{unit} Num_{actv} - \gamma Num_{actv} 
    \end{cases}
\end{equation}
where $\frac{dNum_{unit}}{dt}$ and $\frac{dNum_{actv}}{dt}$ represent the instantaneous growth of the two populations, $t$ is time, $\alpha, \beta, \gamma, \delta$ are positive constants describing the interaction of the killing power of both parties.

\begin{equation}
  \begin{split}
    \frac{dNum_{actv}}{dNum_{unit}} &= \frac{ \delta Num_{unit} Num_{actv} - \gamma Num_{actv}}{\alpha Num_{unit} - \beta Num_{unit} Num_{actv}} \\
        &= \frac{Num_{actv}}{ Num_{unit}} \frac{\delta Num_{unit} - \gamma}{\alpha - \beta Num_{actv}} \\
  \end{split}    
\end{equation}

At this instant, a defender gets more defending resources from the cross-defender collaboration, which forces the attacker to add more resources to maintain possible suppression. Thus, the attacking cost is higher than the previous. More importantly, defenders can trace the attacker and the collateral reflectors based on detected attacking sources and targets in the grid. 

\subsection{The Population of Attacking and Defending Parties}
In our joint-defense area, ISPs, businesses, individuals, and third-party organizations form a defending community, where agents and existing anti-DDoS nodes establish a computing grid. Unlike previous individual strategies in which a user has to combat DDoS of various attacking sources, the grid facilitates establishing a joint front line. From the grid's point of view, the attacking flow may traverse networks of participants. As shown in Fig. 4, a target is now well protected by multiple defending areas. Consequently, such malicious traffic is facing more defending units.

\begin{figure}[t!]
\centering
\includegraphics[width=0.8\columnwidth]{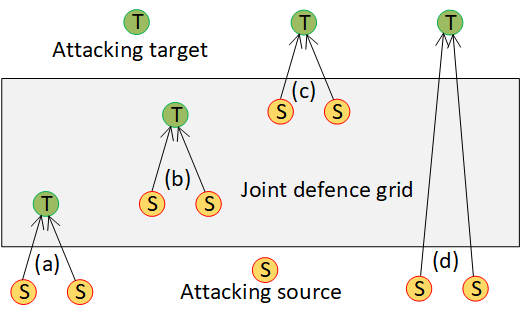}
\caption{Various locations of attackers and defenders.}
\label{f5}
\end{figure}

An attacker can either quit the combat, so the attack fails, or increase the active attacking units until the expense is more than returns.

\begin{equation}
    \frac{\beta Num_{actv} - \alpha}{Num_{actv}} dNum_{actv} + \frac{\delta Num_{unit} - \gamma}{Num_{unit}} dNum_{unit} = 0
\end{equation}

It does not cost much for defenders to add more defending units, as each defender utilizes already-existing infrastructures to fight against attackers in the grid. When neither of the population levels is changing, i.e., 
\begin{equation}
 dNum_{unit} = 0,  dNum_{actv} = 0 
\end{equation}

Combining Equation (8) and (9), we have

\begin{equation}
    \begin{cases}
       (\beta Num_{actv} - \alpha)Num_{unit} &= 0   \\
       (\delta Num_{unit} - \gamma) Num_{actv} &= 0 
    \end{cases}
\end{equation}

The above equation yields two solutions:
\begin{equation}
       Num_{unit} = 0,  Num_{actv} = 0 
\end{equation}
and
\begin{equation}
 Num_{unit} = \frac{\gamma}{\delta}, Num_{actv} = \frac{\alpha}{\beta}
\end{equation}

Equation (11) tells the extinction of both populations. It is an invalid solution unless there is no attacker or defender in the world. Equation (12) represents a fixed level at which both parties sustain their non-zero levels. The two populations oscillate around the rooted value. We used the value of the constant of motion $K$ to represent the closed orbits approximately.

\begin{equation}
    K = Num_{actv}^{\alpha} e^{-\beta Num_{actv}} Num_{unit}^{\gamma} e^{-\delta Num_{unit}}
\end{equation}
The largest value of $K$ is thus attained that meets Equation (12):
\begin{equation}
    K^* = (\frac{\alpha}{\beta e})^{\alpha}(\frac{\gamma}{\delta e})^{\gamma}
\end{equation}
where $K^*$ is the maximum value of the constant.

As demonstrated in Fig. 5, there are four types of attacks involved in the grid, namely, (a) external-to-internal, (b) internal-to-internal, (c) internal-to-external, and (d) external-to-external. When more players join the grid, these types merge and evolve to the internal-to-internal finally. Note that botnets are cheaper because the bots run upon computing nodes of another user's network. When the user joins the grid, the user's network becomes a subnet of the defense grid. In this instance, the system has the potential to eradicate the bot when it attacks any node in the grid. Without enough botnets, the DDoS attack tends to be extremely expensive, which eventually defeats the attackers on the Internet.

\section{The Joint Defense Framework}
Now, we briefly explain the framework and its implementation.

\begin{figure}[t!]
\centering
\includegraphics[width=1\columnwidth]{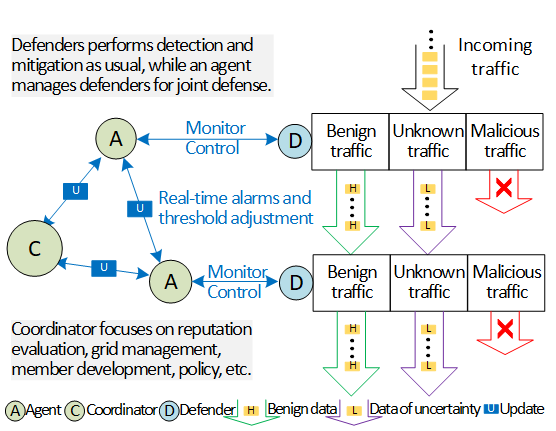}
\caption{The joint-defense framework: defenders perform marking, filtering, rate limiting to incoming traffic as usual. Agents monitor defenders, modify settings, and adjust thresholds to defenders. Agents and coordinators send queries, updates, and alarms to facilitate collaboration.}
\label{f6}
\end{figure}

\subsection{The Framework}
Fig. 6 presents our framework that provides uniform functionality to members in the grid so they can collaboratively fight against DDoS attackers, regardless of the immediate target of whose. The added functions include traffic classifier, rate limiter, and alarm generator. The classifier categorizes incoming data into benign traffic, marked with a higher priority `H'; data of uncertainty, marked with a lower priority `L'; and malicious data to drop.  There are two types of roles in the framework, i.e., agent and coordinator, via which to facilitate the collaboration. Note the primary responsibility of the agent is to share high-level information such as membership, identified bots, legitimacy test results, and knowledge of botnets. Each defender employs individual infrastructure to take practical actions to combat DDoS. The coordinator has the duty for reputation evaluation, membership development, management, policy, etc.

\subsubsection{The Registration and Peer Establishment}

The coordinator invites potential players to join the grid after the player meets our security requirement, as reviewed in Section III. Each invited player installs an agent that carries a valid certificate, issued via human channels when the player joins the grid. When two agents swap high-level information directly, they form a peering relationship. Such a peering relationship reflects the closeness of a pair of upstream and downstream neighbors, where traffic flows on a given path. The downstream neighbor has more accurate knowledge to judge either benign or malicious codes as it is closer to a victim. However, the upstream neighbor has a better position to either prioritize or discard the data. The global view of the peering relationship contributes to the detection and recognition of bots, leading to successful source-end mitigation. In a botnet scenario, such source-end incapacitation signals the killing of bots, resulting in a higher rental price gradually.

In the IoT, both benign traffic and DDoS flows are highly dynamic, so that the peering relationship has to remain updated. Ideally, the relation forms when suspicious flow traverses, and each node removes stale peers if no data of uncertainty pass by after some set timeout. When agents can warranty to deliver collaboration information securely and reliably, the authentication between such agents is of paramount importance. We prevent malicious nodes from joining the grid through a well-designed mutual authentication scheme between agents and the coordinator.  We have detailed the concrete design and implementation of this scheme in \cite{li2021scheme}.

\begin{figure}[t!]
\centering
\includegraphics[width=0.85\columnwidth]{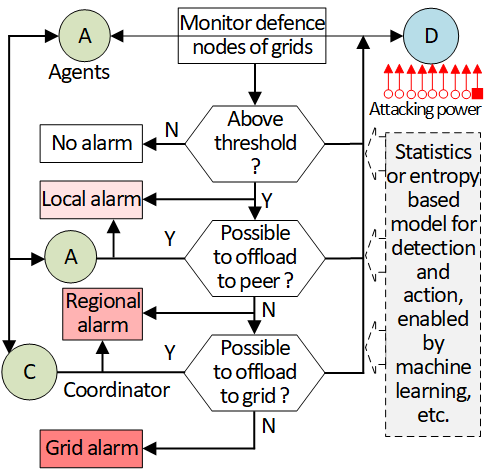}
\caption{The system protection: Levels of alarms.}
\label{f7}
\end{figure}

\subsubsection{The Packet Marking, Policing, and Agent Alarms}

Initially, we proposed higher-priority and lower-priority marking as the protection capacity of individuals may be limited. In a business scenario, if two peers support more classes of traffic separation, they could implement more granular traffic prioritization. For example, the `H' refers to 100\% legitimacy, while businesses can have 95\%, 85\%, 75\% levels of legitimacy to reduce the risks of fault classifications. Without such granularity, the fault (i.e., malicious data marked as benign ones) could only be identified and processed at the victim's end. As studied above, the less mitigation response time, the higher the attacking cost.

According to the various tag of the marked priority, the grid has a different policy to process such data, including bandwidth reservation, reshaping, rate limitation, drop eligibility marking, etc. The purpose is to guarantee the QoE for benign traffic of legitimate users \cite{li2017latency}. The grid should also reserve enough resources for data of uncertainty, which serves new users of legitimacy and opens a window for the grid to keep current knowledge of evolving DDoS attacks. The possible consequence could be the compromise of a part of the grid. At this instant, the alarm is critical to address this issue.

In brief, the alarm is to prevent a network in the grid from being overwhelmed by excessive attacks. Fig. 7 illustrates the alarm system, where agents monitor defense nodes of grids. If the attacking power is higher than the threshold of defense power, the agent raises a local alarm and sends a request to peers for possible offloading. If peers can handle the offloaded attacks, the local alarm keeps. Otherwise, if the attacking power is higher than the joint defense power, they trigger a regional alarm and request the coordinator to allocate more defending units. Now, the agents instruct defending nodes to limit traffic of lower priority. When there is no new unit available to add against the attacking power, they cause a grid alarm. Otherwise, the regional alarm maintains. At this moment, the coordinator and agents instruct defending nodes to kill jobs of lower priority. 

Note the framework can benefit from recent technical advancements, as it allows defenders to deploy preferred solutions in their network. Furthermore, the framework facilitates collaboration between defenders for detection and mitigation, significantly enlarging the defensive power of each participant.

\begin{figure}[t!]
\centering
\includegraphics[width=1\columnwidth]{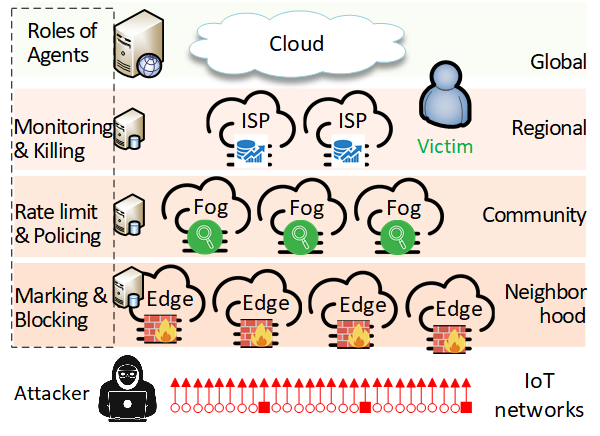}
\caption{The grid implementation.}
\label{f8}
\end{figure}

\subsection{The Implementation: An Edge, Fog, and Cloud Grid}
To further validate our framework, we develop a use case implementation with a grid of edge, fog, and ISP networks \cite{liu2020bayesian}. As shown in Fig. 8, there are four tiers of networks, encompassing neighborhood, community, regional, and global networks \cite{li2018virtualfog}. A neighborhood network can be a business network or an individual network, while a community network refers to a network of fog providers or a campus network. On top of that, the regional network includes networks of local ISPs, and the global network is the network of large ISPs.

The coordinator sits on the top that is responsible for developing memberships, making long-term policies, and performing coordination between participants in the grid through a hierarchy of agents. An attacker can launch distributed attacks from a source either in or out of the grid. The entry point is the edge of the grid where the network starts to combat by blocking or marking the traffic of different priorities. In the worst case, the grid identifies malicious codes at the victim's end. The agent will update upstream agents to acknowledge the source of malicious codes with a combination of the source IP address and the victim's IP. Any attack sourced from a node in the grid is easily blocked and removed at the source-end. When the attacking source is out of the grid, agents request the coordinator to step in for further investigation. Then, the coordinator organizes defending forces to locate the entry point in the grid and taking the corresponding action against such invasion.

Agents may use alarms to notice the level of attacking powers as studied above. No matter in which direction an attack comes from, the victim is behind multiple protection tiers in the grid. As such, the attacker has to invest more attacking forces that come with a higher expense. It is worth noting that a large volume of participants can expedite the maturity of a defending model in a short period, thanks to the sufficient data.

\section{Evaluation}
In this section, we first evaluate the increasing cost of attacks, then implement the framework in our testbed to demonstrate the advantages of joint defense. 
\subsection{Cost Evaluation}
An attacker only needs to defeat one defending system in an individual defense model, while the attacker has to combat all the defending nodes in our joint defense approach. Therefore, the attacker needs to invest more bots to overwhelming a victim. To simplify the calculation, we only consider the botnet expense that an attacker has to pay. According to Equation (5), we need to work out the $Num_{actv}$ and the $PABE$. The active bot refers to the bot staying in the system, causing the consumption of a resource. To this end, the mitigation response time is the determining factor for the per bot price. We have articulated the evaluation in \cite{li2020fleam} and found the attacking cost on the collaborative system is 2.5 times higher than the cost of individual defense on average. We adopt this value in this evaluation.

\begin{table}[t!]
\caption{Attack expense to compromise one object: Individual vs. Joint.}
\centering
\label{T3}
\begin{adjustbox}{max width=1\linewidth}
\begin{tabular}{|c|l|l|l|l|l|l|l|l|l|}
\hline
Individual & \multicolumn{9}{c|}{Joint Defense with Peers of}                                                                                                                                                                              \\ \hline
1          & \multicolumn{1}{c|}{1} & \multicolumn{1}{c|}{2} & \multicolumn{1}{c|}{3} & \multicolumn{1}{c|}{4} & \multicolumn{1}{c|}{5} & \multicolumn{1}{c|}{6} & \multicolumn{1}{c|}{7} & \multicolumn{1}{c|}{8} & \multicolumn{1}{c|}{9} \\ \hline
1          & 7                      & 10.5                   & 14                     & 17.5                   & 21                     & 24.5                   & 28                     & 31.5                   & 35                     \\ \hline
2          & 14                     & 21                     & 28                     & 35                     & 42                     & 49                     & 56                     & 63                     & 70                     \\ \hline
3          & 21                     & 31.5                   & 42                     & 52.5                   & 63                     & 73.5                   & 84                     & 94.5                   & 105                    \\ \hline
\end{tabular}
\end{adjustbox}
\end{table}

\begin{figure}[t!]
\centering
\includegraphics[width=0.9\columnwidth]{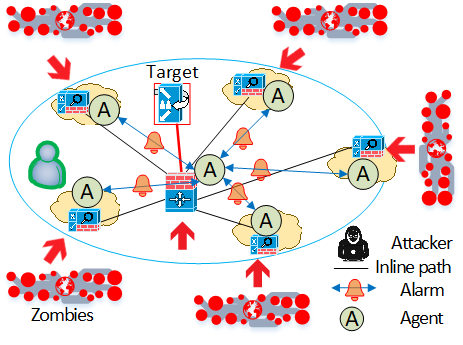}
\caption{The network diagram.}
\label{f9}
\end{figure}

The next factor is the number of active bots required during an attack. The number of bots is proportional to the number of defending units, as investigated in Section IV.B. In our joint defense system, the defense units are $x$ times as which of an individual defense system. Therefore, the botnet expense for an attacker to overwhelm a node in the grid is:

\begin{equation}
  \begin{split}
    Expense_{joint} &= \theta * x * Num_{actv} * PABE \\
               & = \theta * x *  Expense_{botnets} (x=1,2, \cdots)\\ 
  \end{split}    
\end{equation}
where $Expense_{botnets}$ is the expense for a successful compromise in individual defense system.

Table III further numerically compares the expenses. When a defender has one peer, an attacker needs to beat the two defenders to suppress one node in either system. For example, if the expense is \$1 to attack the node in an individual network, the cost is \$7 in the grid. For a defender with peers of 9, the expense is \$35, which is 35 times as previous. The skyrocket of the cost can prevent profit-driven attackers effectively.

\subsection{Experimental Assessment}

\begin{figure*}[hbt!]
\centering
\includegraphics[width=1\linewidth]{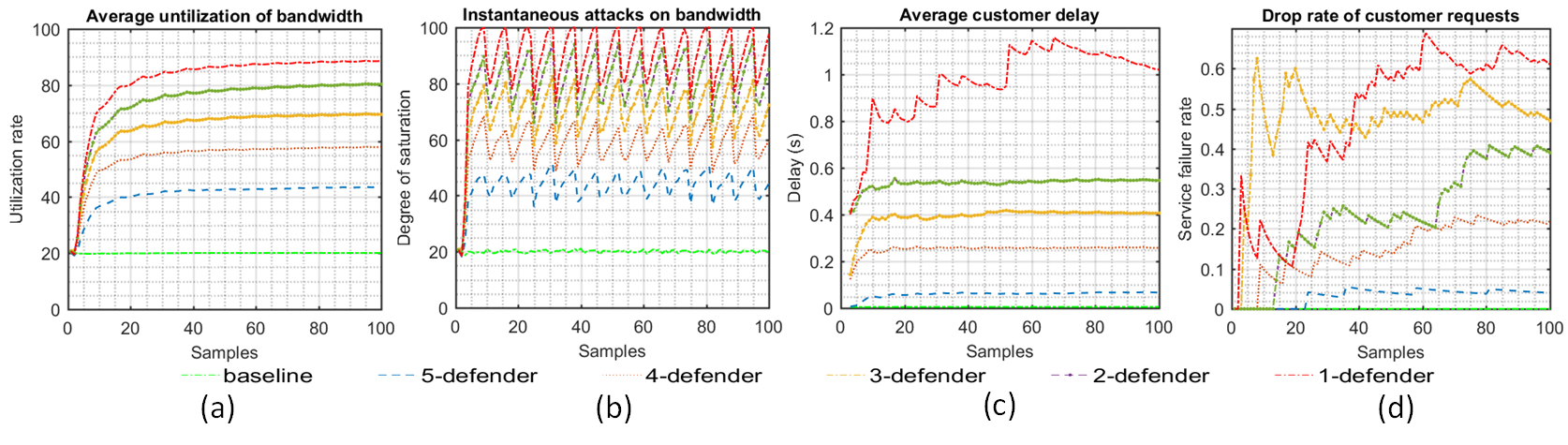}
\caption{The grid implementation.}
\label{f10}
\end{figure*}

\subsubsection{The Settings}
Fig. 9 shows the network diagram of our testbed. The target is a server in a participating defender's network. On the other hand, the defender has five peers that construct a perimeter protecting the target jointly. We use OPNET Modeler \cite{khiat2016study} to implement the testbed, where firewalls are defending nodes, an Ethernet Server is the victim, an Ethernet Workstation is the client, a 1000BaseX Switched LAN is a botnet with 16 bots. 

We use HTTP traffic with Transaction Interarrival Time (TIT) of 10 seconds as the benign traffic, while HTTP with TIT of 0.3 seconds as the malicious flow. The attacker launches an attack every other five minutes, so we could observe many times of attacking and defending progress. We employed 400 bots to suppress the server and the link in the process.

Initially, 400 bots can saturate the red link between the target and the defending node. At this moment, the defender's agent sends alarms with information of IP and victim, requesting help to peers.  Then, the peers limit some traffic towards the victim by adjusting thresholds on inline defending nodes. We observe the bandwidth consumption, delay of legitimate data, and service cancellation rate at the client-end.

\subsection{The Result}
Fig. 10 presents the result. Fig. 10 (a) demonstrates the average utilization of the red link bandwidth. The baseline is about 20\% of utilization that is the state when all defending nodes fight with the 400 bots. The bandwidth consumption rate is approximately 40\% for five peer defenders in the combat. In contrast, the degree of saturation is almost 90\% when the defender fight with bots alone.

Fig. 10 (b) shows the instantaneous bandwidth consumption at different settings of defenders. When the defender combats with bots individually, the attacker can saturate the entire channel at each round of such attacks. Generally, the more defenders join the game, the less bandwidth consumes by bots. To illustrate, four defenders can collaboratively remove about 50\% of malicious traffics.

When the link is saturated, the quality of user experience (QoE) will be much degraded. As illustrated in Fig. 10 (c), the benign traffic delay of legitimate users is around 1 second in the individual defending approach. When more defenders join the combat, the latency for the traffic of legitimate users drops at various levels. In particular, the delay is about 40 milliseconds when five defenders participate in the mitigation.

The link saturation is detrimental to legitimate customers, as they have to cancel their requests due to timeout. Fig. 10 (d) demonstrates the service failure rate is up to above 90\% in the previous individual defense approach. When more defenders start to combat botnets, the drop rate starts to reduce, which is down to 2.5\% with five defenders. Thus, QoE is much better in our joint defense system.

As can be seen, the joint defense approach can handle DDoS threats effectively. With such collaboration, the defense line moves closer to the attacking source through updates between peers. Finally, defenders will be able to destroy bots and catch the attacking agents. In this process, defenders enhance their defense power through collaboration. Our approach takes advantage of existing mitigation infrastructure in the combat, maintaining a lower expense of defense. Conversely, an attacker has to invest more resources with the increment of defending units. As a result, the joint defense system incurs a higher expense, which will cause attackers to lose determination and confidence.

\section{Further Discussion}
Generally, an attacker implements DDoS attacks by flooding an enormous volume of malicious codes to cause resource starvation on the target. A defender can easily detect high-rate attacks but not low-rate attacks (e.g., one packet per hour) due to their stealthy. Unfortunately, by the coordination of millions of zombies, an attacker can easily use botnets to overwhelm a target successfully, even with low-rate attacks. At the victim-side, the defender has the knowledge of such malicious code but lacking control to sourcing nodes. Technically, the attacker employs nodes of one network to attack another defender, vice versa. In consequence, an attacker exploits collaboration across numerous devices and networks for low-cost as well as long-term attacks.

The DDoS attack is one of the most detrimental tools for crackers as it does not need much investment while triggers catastrophic loss to a victim. The main drawback of the traditional solutions against DDoS is that they lack enough consideration to attacking cost, even though the remarkable profit keeps propelling the scale of related DDoS attacks. Only a few papers discussed this point in their works. In \cite{Somani2017cost}, Somani \textit{et al.} studied the cost of mitigation and defense, arguing such expense should always be lower than the business gains received if there was no attack. In \cite{yu2013can}, Yu \textit{et al.} focused on using idle resources in a cloud to enhance server resilience at a low cost. Although the cloud may have sufficient idle resources, the expense to a defender may be less affordable in combat with numerous populations of botnets.

Meanwhile, many scholars worked on reputation-based DDoS mitigation solutions. For example, Dahiya \textit{et al.} assumed that a legitimate user could always follow network rules while a malicious user could not \cite{dahiya2020reputation}. Thus, the cost of an attacker was higher than which of a legitimate user in a reputation-based access system. However, the authors had little consideration of low-rate attacks from pervasive and ubiquitous IoT nodes. These lessons suggest that individual mitigation alone may not be sufficient to combat DDoS attacks in the IoT ecosystem. 

The IoT grid is a group of people and organizations with common interests who work together and communicate more frequently in an Internet community. Our joint framework makes it possible to locate attacking sources within the community by sharing knowledge across defenders with mutual benefits. For the victim, the defender moves the front line closer to the source side, offloading defense duties to another defender and killing the bots at the source-end of the grid edge. At the zombie side, the defender gains an in-depth understanding of low-rate activities and now eradicates these bots before they actively attack the local network.

In comparison to detection, mitigation is easy to implement by filtering and manipulating traffics. Our framework benefits from existing mitigation infrastructure during the combat, which enhances the defending power at a low cost. Meanwhile, we effectuate a much higher expense to an attacker targeting a victim in the grid. Besides, botnets are detrimental to the Internet, as they consume energy and computing power, undermine security pillars, cause a loss in terms of network downtime, finance, and reputation. In this regard, it is necessary for ISP, businesses, individuals, and third-party organizations to join together for the protection of digital assets.

All stakeholders of an IoT grid must work together to combat DDoS attacks. IoT botnets involve myriads of personal devices such as home appliances, web cameras, wearable devices, tablets, etc. These devices are less powerful and easy to be hacked to make up botnets. Due to their volume and closeness to people, they represent enormous attack surfaces. We have achieved a practical solution to address such embarrassments, spanning lightweight protocols, appropriate algorithms, and less-complicated solutions for people with little knowledge of technologies \cite{li2021scheme}.

The joint framework requires secure, reliable, and real-time communication to facilitate collaboration, which may become an extra overhead for simple networks, e.g., a home network. It needs to reduce the communication cost between agents, especially for individuals in the IoT grid. Without the participation of individuals, botnets providers have an endless source for zombies. Consequently, each defender has to face more DDoS attacks. Similarly, it needs to reduce the computational complexity of defenders.

Overall, an attacker will never stop attacking as long as the profit is high enough in a commercial market. The high returns drive the arising of previous, current, and future attackers at every corner of our world. We advocate the joint defense framework to maximize the benefits of closeness between participants in an IoT community, fighting together to beat DDoS by decreasing the attacking profits gradually. As a result, we safeguard our right to Internet availability and data accessibility in the long run. Now, we present Table IV to summarize the works of DDoS studied in this paper.

\begin{table*}[t!]
\caption{A summary of works studied in this paper.}
\centering
\label{T4}
\begin{adjustbox}{max width=1\linewidth}
\begin{tabular}{|c|c|c|c|c|c|c|}
\hline
\multirow{2}{*}{\begin{tabular}[c]{@{}c@{}}Reference\\ Number\end{tabular}} & \multirow{2}{*}{\begin{tabular}[c]{@{}c@{}}Year\\ of the\\ Work\end{tabular}} & \multicolumn{5}{c|}{Factors Considered in  the Combat against DDoS}                                                                                                                                                                                                                                                                        \\ \cline{3-7} 
                                                                            &                                                                               & Detection and Mitigation Technology        & \begin{tabular}[c]{@{}c@{}}Attack \\ Technology\end{tabular} & \begin{tabular}[c]{@{}c@{}}Mitigation \\ Cost Studied\end{tabular} & \begin{tabular}[c]{@{}c@{}}Attacking \\ Cost Studied\end{tabular} & Participants                                                                          \\ \hline
{[}24{]}                                                                    & 2014                                                                          & Cloud, resource scaling, queueing theory   & Generic                                                      & Yes                                                                & No                                                                & Businesses                                                                            \\ \hline
{[}19{]}                                                                    & 2016                                                                          & Proxy-based attacker isolation             & Generic                                                      & No                                                                 & No                                                                & Businesses                                                                            \\ \hline
{[}23{]}                                                                    & 2016                                                                          & Resource scaling, splittable units         & Generic                                                      & No                                                                 & No                                                                & Businesses                                                                            \\ \hline
{[}4{]}                                                                     & 2017                                                                          & Focused on botnets-based attacks           & Botnets                                                      & No                                                                 & No                                                                & Not available                                                                         \\ \hline
{[}6{]}                                                                     & 2017                                                                          & Focused on attacker's profits              & Botnets                                                      & No                                                                 & Yes                                                               & Not available                                                                         \\ \hline
{[}7{]}                                                                     & 2017                                                                          & Resource management, cloud, auto scaling   & Generic                                                      & Yes                                                                & No                                                                & Businesses                                                                            \\ \hline
{[}8{]}                                                                     & 2017                                                                          & Filtering-based detection and mitigation   & Generic                                                      & No                                                                 & No                                                                & Businesses                                                                            \\ \hline
{[}9{]}                                                                     & 2017                                                                          & SDN, collaborative anomaly detection       & Generic                                                      & No                                                                 & No                                                                & ISP and customers                                                                     \\ \hline
{[}18{]}                                                                    & 2017                                                                          & SDN, proactive defense                     & Generic                                                      & No                                                                 & No                                                                & Businesses                                                                            \\ \hline
{[}12{]}                                                                    & 2018                                                                          & SDN, multi-level mitigation                & Generic                                                      & No                                                                 & No                                                                & Businesses                                                                            \\ \hline
{[}16{]}                                                                    & 2018                                                                          & Big data, machine learning                 & Generic                                                      & No                                                                 & No                                                                & Businesses                                                                            \\ \hline
{[}21{]}                                                                    & 2018                                                                          & Blockchain                                 & Generic                                                      & No                                                                 & No                                                                & Businesses                                                                            \\ \hline
{[}22{]}                                                                    & 2018                                                                          & Machine learning, neural networks          & Generic                                                      & No                                                                 & No                                                                & Businesses                                                                            \\ \hline
{[}10{]}                                                                    & 2019                                                                          & SDN, cloud, flow table scaling             & Generic                                                      & No                                                                 & No                                                                & Businesses                                                                            \\ \hline
{[}11{]}                                                                    & 2019                                                                          & Blockchain, SDN                            & Botnets                                                      & No                                                                 & No                                                                & Businesses                                                                            \\ \hline
{[}17{]}                                                                    & 2019                                                                          & Fog, cloud, NFV                            & Botnets                                                      & No                                                                 & No                                                                & Businesses                                                                            \\ \hline
{[}5{]}                                                                     & 2020                                                                          & Focused on DDoS scale                      & Botnets                                                      & No                                                                 & No                                                                & Not available                                                                         \\ \hline
{[}13{]}                                                                    & 2020                                                                          & Deep learning                              & Botnets                                                      & No                                                                 & No                                                                & ISP                                                                                   \\ \hline
{[}14{]}                                                                    & 2020                                                                          & Reinforced learning, host-based mitigation & Generic                                                      & No                                                                 & No                                                                & Businesses                                                                            \\ \hline
{[}15{]}                                                                    & 2020                                                                          & Federated learning, cost-based mitigation  & Generic                                                      & No                                                                 & Yes                                                               & Businesses                                                                            \\ \hline
{[}20{]}                                                                    & 2020                                                                          & Reputation-based mitigation                & Generic                                                      & No                                                                 & No                                                                & Businesses                                                                            \\ \hline
{[}28{]}                                                                    & 2020                                                                          & Task offloading, machine learning          & Generic                                                      & No                                                                 & No                                                                & Businesses                                                                            \\ \hline
{[}31{]}                                                                    & 2020                                                                          & Reputation-based, price strategy           & Generic                                                      & No                                                                 & Yes                                                               & Businesses                                                                            \\ \hline
This Work                                                                   & 2020                                                                          & Cost-based, joint-defense, all others      & Generic, botnets                                                      & Yes                                                                & Yes                                                               & \begin{tabular}[c]{@{}c@{}}ISPs, third-parties,\\ businesses, individuals\end{tabular} \\ \hline
\end{tabular}
\end{adjustbox}
\end{table*}

\section{Conclusion and Future Work}
Botnets enable attackers to attack a victim at a lost cost, rewarding them enormous returns financially. An attacker can install bots in one network, from which to strike victims of another defender. Without enough collaboration between defenders, it is critically difficult to eradicate such rooted zombies in the IoT domain. Therefore, we investigate joint defense where many defenders collaboratively combat with one attacker in an IoT community, resulting in an agent-coordinator based framework to mitigate DDoS. At the victim-end, the agent informs the coordinator and peer agents of attacking sources and the victim's information. Accordingly, agents instruct defending nodes to kill malicious codes at the source end in real-time.

The joint approach forces an attacker to employ more botnet populations during an attack, incurring much higher expense and thus preventing profit-driven crackers. We furnish proofs of the incurred cost growing linearly with the increment of participants. The skyrocket of attacking levy, in turn, effectively stops and stifles the attackers' attempts, ultimately defeats them, and protects our digital assets in the IoT community. In the future, we will work on solutions to trace malicious codes sourced from out of the IoT grid, increasing their exposures so that the criminals have nowhere to hide.

\bibliographystyle{IEEEtran}
\bibliography{ddos}

\end{document}